\documentstyle[preprint,aps,epsf,rotate]{revtex}
\begin{document}
\draft
%
\title{Stochastic Loewner Evolution and Dyson's Circular Ensembles}
\author{John Cardy}
\address{Institute for Advanced Study, Princeton NJ 08540 \\
 and University of Oxford, Department of Physics -- Theoretical
         Physics, 1 Keble Road, Oxford OX1 3NP, U.K.\footnote{Address for
correspondence} \\
         and All Souls College, Oxford.}
%
%
\maketitle
\begin{abstract}
Stochastic Loewner Evolution (SLE$_\kappa$) has been introduced as a 
description of 
the continuum limit of cluster boundaries in two-dimensional critical systems. 
We show that the problem of $N$ radial SLEs in the unit disc
is equivalent to Dyson's Brownian motion on the boundary of the disc, with 
parameter $\beta=4/\kappa$. As a result various equilibrium critical models give
realisations of circular ensembles with $\beta$ different from the
classical values of $1,2$ and $4$ which correspond to symmetry classes of
random U$(N)$ matrices.
Some of the bulk critical exponents are related to the spectrum of the associated
Calogero-Sutherland hamiltonian.
The main result is also checked against the predictions of conformal field theory.
{\bf An erratum to the published version has been added in which the
above comparison is made more explicit.}
\end{abstract}
%
%
\subsection*{Introduction}
Recently a new method for understanding the scaling limit of conformally invariant
two-dimensional critical systems has been introduced by Schramm\cite{S} and 
developed by Lawler, Schramm and Werner\cite{LSW} (LSW).
This is known as stochastic Loewner evolution (SLE). It relies on the fact that 
many such systems may be realised geometrically in terms of sets of random curves, 
whose statistics can be described by a stochastic dynamical process. SLE is the
continuum limit of this process.

There is in fact a continuous family of SLE processes, labelled by a real parameter
$\kappa\geq0$. Different values of $\kappa$ are supposed to correspond to different
universality classes  of critical phenomena.
For example, for $4\leq\kappa\leq8$ they describe the perimeters
of the Fortuin-Kastelyn clusters of the $Q$-state Potts model with $4\geq Q\geq0$,
while for $2\leq\kappa\leq4$ they describe the graphs of the high-temperature
expansion of the O$(n)$ model with $-2\leq n\leq2$ (dual to the boundaries
of critical Ising spin clusters for $n=1$), 
as well as the external boundaries of F-K clusters.
This correspondence has so far been proven 
rigorously only in a few cases\cite{proof}. 
However, if the continuum limit of the lattice curves exists and is
conformally invariant, it must be described by SLE\cite{LSW}. 

Under these assumptions, LSW\cite{LSW}
have rederived many of the known
results for two-dimensional critical behaviour which have been found by
less rigorous approaches such as Coulomb gas methods and conformal field theory, as
well as establishing some new ones. One aspect of the connection with conformal
field theory has 
recently been pointed out by Bauer and Bernard\cite{BB} 
and Friedrich and Werner\cite{FW}.

The particular setting we consider in this note is as follows: consider a critical
system in a disc of radius $R$, with a puncture at the origin of radius 
$\epsilon$, in the limit when $R$ is much larger than $\epsilon$ and the
lattice spacing $a$. Suppose there are exactly $N$ open curves connecting the inner and
outer boundaries. In addition, there are no open curves which begin and end on the
outer boundary. See Fig.~1.
For example, these curves could be mutually avoiding self-avoiding 
walks\cite{LSWSAW},
or the external boundaries of percolation clusters (both $\kappa=\frac83$), or
the boundaries of critical Ising spin clusters ($\kappa=3$.) (In these last two
cases we assume that the ensemble is conditioned so as to satisfy the above.)
Another example, conjectured to correspond to to $\kappa=4$, is when the curves are the 
level lines of a two-dimensional crystalline surface
at the roughening transition, and there is a 
screw dislocation of strength $N$ located at the origin.
As long as $\kappa\leq4$ it is known\cite{SR}
that, in the continuum limit, these curves are simple, that is, they
self-intersect with probability zero. For the same reason, the positions at
which the curves intersect the outer boundary, labelled by complex numbers
$Re^{i\theta_j}$, are well-defined for $\kappa\leq4$. 

Our main result is that, for $R\gg\epsilon$, the joint probability density  
function (p.d.f.) of these points is given by Dyson's circular 
ensemble\cite{dyson1}
\begin{equation}
\label{pdf}
P_{\rm eq}(\{\theta_j\})
\propto\prod_{1\leq j<k\leq N}\Big|e^{i\theta_j}-e^{i\theta_k}
\Big|^\beta
\end{equation}
with $\beta=4/\kappa$. 
Our argument proceeds by showing that the SLE process appropriate for this
situation contains the Brownian process invented by Dyson\cite{dyson2}, 
whose equilibrium
distribution is given by (\ref{pdf}), with time being asymptotically 
proportional to $\ln(R/\epsilon)$.  

The distribution (\ref{pdf}) is known to describe the statistics of the eigenvalues
of random unitary matrices in the orthogonal, unitary and symplectic ensembles
for $\beta=1,2$ and $4$ respectively. Our arguments thus provide simple physical
realisations of this ensemble for other values of $\beta>1$, for example 
$\beta=\frac32$ (self-avoiding walks) and $\beta=\frac43$ (Ising spin cluster
boundaries.) 

We also check (\ref{pdf}) against the predictions of conformal field theory (CFT).
There is a subtle factor of $\frac12$ in the exponent which we elucidate.
Dyson's process is known to be related by a similarity transformation to the 
quantum Calogero-Sutherland model\cite{CS}. 
We point an interesting connection between the dilatation operator 
$D\equiv L_0+\overline L_0$ of CFT and the Calogero-Sutherland hamiltonian.

It turns out that the eigenvalues
of this hamiltonian may, with suitable boundary conditions, correspond to bulk scaling
exponents of the models with $\kappa>4$, an example being the one-arm exponent
computed by LSW\cite{1arm}.

\subsection*{Multiple SLEs}

Our arguments are based on an $N$-particle generalisation of radial SLE.
A single radial SLE describes the continuum limit of a curve in the unit disc
$U:=\{z:|z|<1\}$ which begins on the boundary at time $t=0$ and ends up at the
origin $0$ as $t\to\infty$. Let $K(t)$ be the hull of the process up to time
$t$ (for $\kappa\leq4$ this is just the set of points on the curve.) There
is a conformal mapping $g_t:U\setminus K(t)\to U$, 
such that $g_t(0)=0$ and $g_t'(0)>0$.  LSW argue that $g_t(z)$
may be chosen so as to satisfy an evolution equation
\begin{equation}
\label{rsle}
\dot g_t(z)=-g_t(z)\,{g_t(z)+e^{i\sqrt\kappa B(t)}\over
g_t(z)-e^{i\sqrt\kappa B(t)}}
\end{equation}
where $B(t)$ is a standard one-dimensional Brownian process with ${\bf E}[B(t)^2]=t$. 
Note that time has been reparametrised so that $g_t'(0)=e^t$. 
(\ref{rsle}) is the standard form of radial SLE, which maps the trace of the SLE
into the point $e^{i\sqrt\kappa B(t)}$ on the boundary, but for
our purposes it is more convenient to consider
$\hat g_t(z;\theta)\equiv g_t(z)e^{i(\theta-\sqrt\kappa B(t))}$, which
maps the trace into $e^{i\theta}$, and satisfies\footnote{In the Ito convention there is 
an additional term $-\frac\kappa2\hat gdt$ on the right hand side. This disappears again
after making the global rotation leading from (\ref{GNprime}) to (\ref{GN}).}
\begin{equation}
\label{rsle2}
d\hat g_t(z;\theta)=-\hat g_t(z;\theta)\,{\hat g_t(z;\theta)+
e^{i\theta}\over\hat g_t(z;\theta)-e^{i\theta}}\,dt
-i\hat g_t(z;\theta) \sqrt\kappa\, dB(t)
\end{equation}

Now consider $N$ SLE's which start from distinct points $\{e^{i\theta_j}\}$ on the
boundary, with $1\leq j\leq N$. Let $K_j(t)$ be the hull of the $j$th SLE. For
$\kappa\leq4$ these are segments of non-intersecting simple curves.
Let $G^{(N)}_t(z)$ be a function which conformally maps
$U\setminus\cup_{j=1}^NK_j(t)$ onto $U$, with $G^{(N)}_t(0)=0$ and 
${G^{(N)}_t}'(0)>0$.
Then we shall argue that $G^{(N)}_t(z)$ may be chosen to satisfy
\begin{equation}
\label{GN}
{\dot G}^{(N)}_t=-G^{(N)}_t\sum_{j=1}^N{G^{(N)}_t+e^{i\theta_j(t)}\over
G^{(N)}_t-e^{i\theta_j(t)}}
\end{equation}
where
\begin{equation}
\label{dyson}
d\theta_j(t)=\sum_{k\not=j}\cot((\theta_j(t)-\theta_k(t))/2)\,dt
+\sqrt\kappa\, dB_j(t)
\end{equation}
and $B_j(t)$ are $N$ independent Brownian motions, starting at the origin.

To see this, consider the infinitesimal transformation
$G^{(N)}_{t+dt}\circ (G^{(N)}_t)^{-1}$
and note that this may be obtained by allowing each
SLE to evolve independently according to (\ref{rsle2}) over a time $dt$:
\begin{equation}
G^{(N)}_{t+dt}\circ (G^{(N)}_t)^{-1}=
g^{(N)}_{dt}(\theta_N(t))\circ g^{(N-1)}_{dt}(\theta_{N-1}(t))
\circ\ldots\circ g^{(1)}_{dt}(\theta_1(t))
\end{equation}
During the evolution of the $j$th SLE, $G^{(N)}_t$ evolves according to
(\ref{rsle2}), with $\theta=\theta_j(t)$ and $B(t)=B_j(t)$, but the other
$\theta_k(t)$ with $k\not=j$ also evolve according to 
\begin{equation}
d\theta_k(t)=i\,{e^{i\theta_k(t)}+e^{i\theta_j(t)}\over 
e^{i\theta_k(t)}-e^{i\theta_j(t)}}\,dt
-\sqrt\kappa\, dB_j(t)
=\cot((\theta_k(t)-\theta_j(t))/2)dt-\sqrt\kappa\, dB_j(t)
\end{equation}
Thus, after evolving every SLE with $j=1,\ldots,N$, we have 
\begin{equation}
\label{GNprime}
dG^{(N)}_t=-G^{(N)}_t\sum_{j=1}^N{G^{(N)}_t+e^{i\theta_j(t)}\over
G^{(N)}_t-e^{i\theta_j(t)}}\, dt-iG^{(N)}_t\sqrt\kappa\sum_{j=1}^NdB_j(t)
\end{equation}
where
\begin{equation}
d\theta_j(t)=\sum_{k\not=j}\cot((\theta_j(t)-\theta_k(t))/2)dt
-\sqrt\kappa\sum_{k\not=j}dB_k(t)
\end{equation}
It is now simpler to rotate the whole disc through an angle 
$\sqrt\kappa\sum_{j=1}^NdB_j(t)$, after which we obtain (\ref{GN}) and
(\ref{dyson}) as claimed.

Eq.~(\ref{dyson}) is the Dyson process\cite{dyson2}. It may be written as
\begin{equation}
d\theta_j=-{\partial V\over\partial\theta_j}\,dt+\sqrt\kappa\,dB_j(t)
\end{equation}
where $V\equiv-2\sum_{j<k}\ln|\sin((\theta_j-\theta_k)/2)|$.
At late times, the distribution of the $\{\theta_j(t)\}$
tends towards an equilibrium p.d.f. at temperature $\kappa/2$:
\begin{equation}
P_{\rm eq}(\theta_1,\ldots,\theta_N)\propto
e^{-2V/\kappa}=\prod_{j<k}|\sin((\theta_j-\theta_k)/2)|^{4/\kappa}
\propto \prod_{j<k}|e^{i\theta_j}-e^{i\theta_k}|^\beta
\end{equation}
This is Dyson's circular ensemble\cite{dyson1}, with
\begin{equation}
\beta=4/\kappa\quad.
\end{equation}

So far, the starting points $\{\theta_j\}$ have been taken to be fixed.
Now consider an ensemble of these, generated by the configurations of some
bulk critical ensemble in the interior of $U$ (conditioned if necessary 
on the existence of exactly $N$ curves.)
We argue that if this corresponds to a conformally invariant bulk system,
the p.d.f. of the $\{\theta_j\}$ must be given by $P_{\rm eq}$.
Let $K(\{\theta_j\},t)$ be the union of the
$N$ hulls up to time $t$, given that they start at the points $\{e^{i\theta_j}\}$.
The expectation value of any observable may be
taken by first conditioning on the subset $K(\{\theta_j\},t')$, with $t'<t$. 
By conformal invariance, the distribution of 
$K(\{\theta_j\},t)\setminus K(\{\theta_j\},t')$ is the same as that of its image
under $G^{(N)}_{t'}$, namely $K(\{\theta_j(t')\},t-t')$. Averaging over
$K(\{\theta_j\},t')$ is equivalent to averaging over the $\{B_j\}$ up to 
time $t'$. 
Taking $t\to\infty$, we conclude that 
\begin{equation}
P(\{\theta_j\})={\bf E}_{\{B_j(t''):t''\in[0,t']\}}\Big[P(\{\theta_j(t')\})\Big]
\end{equation}
that is
$P(\{\theta_j\})$ is stationary under the process (\ref{dyson}), and 
must therefore be equal to $P_{\rm eq}$.

This is strictly valid only when the SLEs are allowed to reach the origin.
To discuss the case when they reach only the circle $|z|=\epsilon$,
it is helpful to map conformally
the annulus to a cylinder of length  $\ell\equiv\ln(R/\epsilon)$. 
The points $e^{i\theta_j}$ are now arrayed around one end of the cylinder. 
Since ${\dot G}^{(N)}_t=NG^{(N)}_t(1+O(G^{(N)}_t))$ as $G^{(N)}_t\to0$, we see
that as long as $\ell\gg1$ the effect of the evolution is to reduce the
length of the cylinder at a rate $\dot\ell=-N$. Meanwhile the points $\{\theta_j\}$
are moving according to (\ref{dyson}). 
The approach of their distribution to equilibrium
is expected to be exponentially fast with a rate constant $O(1)$.
Thus, as long as $\ell\gg N$, we may apply the same argument as above, and deduce that
the distribution of the $\{\theta_j\}$ is given by $P_{\rm eq}$, with corrections
suppressed by powers of $\epsilon/R$. The same should
apply on a lattice, as long as the spacing $a<\epsilon\ll R$. 

\subsection*{Comparison with conformal field theory}

The crucial assumption of conformal invariance made in deriving the above result 
would appear to be stronger than the analogous statement for $N=1$.
In particular, it is not clear from this point of view why invariance under 
the uniformising transformation $G^{(N)}_t$, which assumes that the curves grow 
at the same rate, is to be chosen among other possibilities, although it appears
to be the most natural one. 
For this reason we have checked our main result (\ref{pdf}) using methods of CFT.
In this language we expect the joint p.d.f. to be given by the correlation function
in the $O(n)$ conformal field theory
\begin{equation}
\label{correlator}
\langle\Phi_{a_1\ldots a_N}(0)
\phi_{a_1}(Re^{i\theta_1})\ldots\phi_{a_N}(Re^{i\theta_N})\rangle
\end{equation}
where $\phi_a(Re^{i\theta})$ is a boundary $1$-leg operator carrying $O(n)$ 
index $a$, and $\Phi$ is a bulk $N$-leg operator.
By choosing the $a_j$ to be all different, we ensure that the curves all reach the
origin without annihilating. In \cite{JCsurf} it was conjectured that the operators
$\phi_a$ correspond to Virasoro representations labelled by $(1,2)$ in the Kac
classification. These have a null state at level 2,
and therefore their correlators satisfy second-order linear partial differential
equations with respect to each of the $\theta_j$ (the BPZ equations\cite{BPZ}). 
The general solution for such an $N+1$-point correlator is not however known. 
Instead, we may take the form (\ref{pdf}) as an ansatz, and check whether it 
satisfies these equations. Even this is somewhat tedious, and we have carried it through
only for $N=2$. Alternatively, 
one may check whether (\ref{pdf}) satisfies the fusion rules 
which follow from the BPZ equations. These determine the behaviour of the correlator
(\ref{correlator}) in the limits when (say) $p$ of the $\theta_j$ approach each other.
Suppose, for example, that $|\theta_j-\theta_k|=O(\delta)$ for $1\leq j\leq p$
and $1\leq k\leq p$, with $2\leq p\leq N$. In the limit $\delta\to0$ we may use
the operator product expansion (OPE)
\begin{equation}
\label{ope}
\prod_{j=1}^p\phi_{a_j}(\theta_j)\propto \delta^{x_p-px_1}\,
\phi_{a_1\ldots a_p}(\theta_1)
\end{equation}
where $\phi_{a_1\ldots a_p}$ is the boundary $p$-leg operator, and $x_p$ is its
scaling dimension. Given that the 1-leg operator corresponds to $(1,2)$, the
fusion rules determine the allowed values of $x_p$ which may occur on the right hand side
of (\ref{ope}): they are the scaling dimensions $h_{1,n+1}$ of the $(1,n+1)$
operators, with $0\leq n\leq p$ and $p-n$ even. Duplantier and Saleur\cite{DupSal}
argued that the $p$-leg operator must in fact correspond to $n=p$.
Using the Kac formula $h_{1,p+1}=p(2p+4-\kappa)/2\kappa$ then gives the exponent
in (\ref{ope}) to be simply $p(p-1)/\kappa$.

On the other hand, we may simply take the appropriate limit in the ansatz 
(\ref{pdf}) to find a dependence on $\delta$ of the form 
$\prod_{1\leq j<k\leq p}\delta^\beta=\delta^{p(p-1)\beta/2}$.
Comparing these two expressions apparently gives $\beta=2/\kappa$, not 
$4/\kappa$ as found above. One may take further short-distance limits 
within (\ref{pdf}): the results all consistent with Duplantier and Saleur\cite{DupSal}
but only if $\beta=2/\kappa$.\footnote{We remark that the reason that the
$N+1$-point correlator (\ref{correlator}) has such a simple form is that there is
only ever a single term on the right-hand side of OPEs like (\ref{ope}). This is
because all the $O(n)$ indices on the left-hand side are different, and therefore
the operators on the right hand side transform according to the totally symmetric 
representation of $O(n)$. The other possible terms, with $n<p$, correspond to
other representations, and could only arise if some of the $a_j$ were equal.}

The resolution of this paradox is as follows: the correlator (\ref{pdf}) may be written
in operator language as 
\begin{equation}
\langle\Phi_N|e^{-tD}|\{\theta_j\}\rangle
\end{equation}
where $t=\ln(R/\epsilon)$, $D\equiv L_0+\overline L_0$ is the generator of scale 
transformations, and $|\{\theta_j\}\rangle$ is a boundary state.
The usual formalism of CFT assumes that $D$ is self-adjoint: this originates
in the invariance of the bulk theory under inversions $z\to-1/z$. 
Equivalently, on the cylinder, $e^{-aD}$ is the continuum limit of the transfer
matrix, which, for many lattice models, may also be chosen to be self-adjoint. But for the loop representation of the $O(n)$ model this is not the case: the only practicable 
transfer matrix which has been employed\cite{transfer}
acts, at a given time $t$, on the space spanned by a basis defined
by the positions of the points at which the loops intersect the chosen time slice, 
together with their relative connections in the `past', but not the `future'. This 
asymmetry leads to a transfer matrix $T$ which is not self-adjoint.
In our case, the points at which the loops intersect a given time slice can either be
connected back to the points $e^{i\theta_j}$ at $t=0$, or to each other via loops which 
close in the past at times $t>0$ (see Fig.~2). Let $\Pi$ be the projection
operator in the full space in which $T$ acts, onto the subspace spanned by the
boundary states $|\{\theta_j\}\rangle$. That is, $\Pi$ traces over the positions of the
points not connected to $t=0$.
Then $\widetilde T\equiv \Pi T$ acts wholly within this
subspace. Note that $\int\prod_jd\theta_j\langle\{\theta_j\}|$ is a left eigenstate of
$\widetilde T$ with unit eigenvalue.

As usual, we may also think of $\widetilde T$ as acting on $L^2$ functions of
the $\{\theta_j\}$ through
\begin{equation}
(\widetilde Tf)(\{\theta_j\})=\int\prod_jd\theta_j'
\langle\{\theta_j\}|\widetilde T|\{\theta'_j\}\rangle f(\{\theta_j'\})
\end{equation}
so that ${\widetilde T}^{\dag}{\bf 1}=1$. 
Moreover, as $\ln(R/\epsilon)\to\infty$, the joint p.d.f. of the 
$\{\theta_j\}$ is given by the right eigenfunction of $\widetilde T$ with 
the largest eigenvalue.

On the other hand,
the Langevin equation (\ref{dyson}) yields the Fokker-Planck equation 
$\dot P={\cal L}P$ for the evolution
of $P(\{\theta_j\},t)$ where
\begin{equation}
{\cal L}=\sum_{j=1}^N{\partial\over\partial\theta_j}
{\partial V\over\partial\theta_j}
+{\kappa\over 2}{\partial^2\over\partial\theta_j^2}
\end{equation}
where ${\cal L}P_{\rm eq}=0$ and ${\cal L}^{\dag}{\bf 1}=0$. 
Now the conformal mapping $G_t^{(N)}\propto e^{Nt}$ acts as a scale transformation
near the origin,
while its action on the unit disc is described by $\cal L$.
We therefore conjecture that this is nothing but the continuous version
of $\widetilde T$, more precisely, $\Pi T^t\sim e^{at{\cal L}/N}$, 
acting on the subspace.

Now for any dynamics which satisfies detailed balance $\cal L$ is related
to a self-adjoint operator $H$ by a similarity transformation:
\begin{equation}
H=-P_{\rm eq}^{-1/2}{\cal L}P_{\rm eq}^{1/2}
\end{equation}
The ground state eigenfunction of $H$ is then $P_{\rm eq}^{1/2}$. 
This square root
is the origin of the discrepancy between the CFT result and (\ref{pdf})
with $\beta=4/\kappa$: the self-adjoint operator $e^{-Ht}$ 
is proportional to the scale transformation operator $e^{-Dt}$
where $D=L_0+\overline L_0$ of CFT, acting on the subspace spanned by the 
boundary states, that is $e^{-Ht/N}\propto\Pi e^{-Dt}\Pi^{\dag}$.
But it does not give the continuum limit of the correct transfer matrix.
By assuming that this limit was self-adjoint, which is assumed in the standard 
formulation of CFT, we found, erroneously, the square root of the
correct result $(\ref{pdf})$ with $\beta=4/\kappa$.

In the case of Dyson's brownian motion the hamiltonian $H$ is that of the
quantum Calogero-Sutherland model\cite{CS} 
\begin{equation}
\label{CS}
{\cal H}=-\frac\kappa 2\sum_j{\partial^2\over\partial\theta_j^2}
+{2-\kappa\over2\kappa}\sum_{j<k}{1\over\sin^2(\theta_j-\theta_k)/2}
-{N(N-1)\over2\kappa}
\end{equation}

It is interesting to note that the adjoint operator
${\cal L}^{\dag}=e^{2V/\kappa}{\cal L}e^{-2V/\kappa}$
has the form
\begin{equation}
\label{Ldag}
{\cal L}^{\dag}=\frac\kappa 2\sum_j{\partial^2\over\partial\theta_j^2}
+\sum_j\sum_{k\not=j}\cot((\theta_j-\theta_k)/2){\partial\over\partial\theta_j}
\end{equation}
${\cal L}^{\dag}$ is the generator for a typical first-passage problem. For
example, in the case $N=2$, the probability $h(\theta_1,\theta_2;t)$
that the two particles have not
met up to time $t$, given that they started from $(\theta_1,\theta_2)$
satisfies $\partial_th={\cal L}^{\dag}h$. In fact, with
$\theta\equiv\theta_1-\theta_2$, and rescaling $2t\to t$, this is just the
equation derived by LSW\cite{1arm}, whose lowest non-trivial eigenvalue
gives the one-arm exponent, related to the fractal dimension of F-K clusters
for $\kappa=6$. Eigenfunctions of ${\cal L}^{\dag}$ behave near $\theta=0$
as $\theta^\alpha$ where $\alpha=0$ or $1-4/\kappa$. When $\kappa\leq4$ the
appropriate solution corresponds to $\alpha=0$, or $h=1$, consistent with
the result that SLE is a simple curve, but when $\kappa>4$ the solution
is non-trivial. LSW\cite{1arm} argue that the appropriate boundary
condition at $\theta=2\pi$ for the one-arm problem
is $\partial h/\partial\theta=0$, and that the solution is then
$\big(\sin(\theta/4)\big)^{1-4/\kappa}\,e^{-\lambda t}$
with $\lambda=(\kappa^2-16)/32\kappa$. This is of course also an eigenvalue
of the Calogero-Sutherland hamiltonian (\ref{CS}), and it raises the question
as to whether other bulk scaling dimensions are given by 
eigenvalues of Calogero-Sutherland systems with suitable boundary conditions.

\noindent{\it Acknowledgements:}
The author would like to thank M.~Aizenman, D.~Bernard, B.~Nienhuis,
S.~Smirnov and W.~Werner
for critical comments and suggestions on an earlier version of this paper.
This work was carried out while the author was a member of the Institute for
Advanced Study.
He thanks the School of Mathematics and the School of Natural Sciences
for their hospitality.
This stay was supported by the Bell Fund, the James D.~Wolfensohn Fund,
and a grant in aid from the Funds for Natural Sciences.

\noindent{\bf Erratum to the published version.}

In the published paper\cite{JC} (as its appears above), 
it was noted that the only form of the result for the
joint p.d.f. of the boundary points that is consistent with conformal
field theory (CFT) is
$\prod_{j<k}(e^{i\theta_j}-e^{\theta_k})^{2/\kappa}$, which is different
from the equilibrium distribution of the corresponding Dyson process,
where the exponent is $4/\kappa$. A possible explanation of this
discrepancy was given. 

Recently\cite{JCCS}, however, we have performed an \em ab initio \em CFT
calculation. This confirms the exponent $2/\kappa$, but it also shows the
correct source of the discrepancy lies in the assumption above Eq.~(13)
that the measure on the curves is conformally invariant. 
This is too strong -- if instead we allow for it to be invariant up to
a conformal factor 
$\prod_j|g_t'(e^{i\theta_j})|^{h_{2,1}}$ (where
$h_{2,1}=(6-\kappa)/2\kappa$), the results of the two computations
agree. 

The corresponding Calogero-Sutherland model then turns out out to have
$\beta=8/\kappa$. The reader is referred to \cite{JCCS} for details.

%
%
\begin{figure}
\label{fig1}
\centerline{
\epsfxsize=7cm
\epsfbox{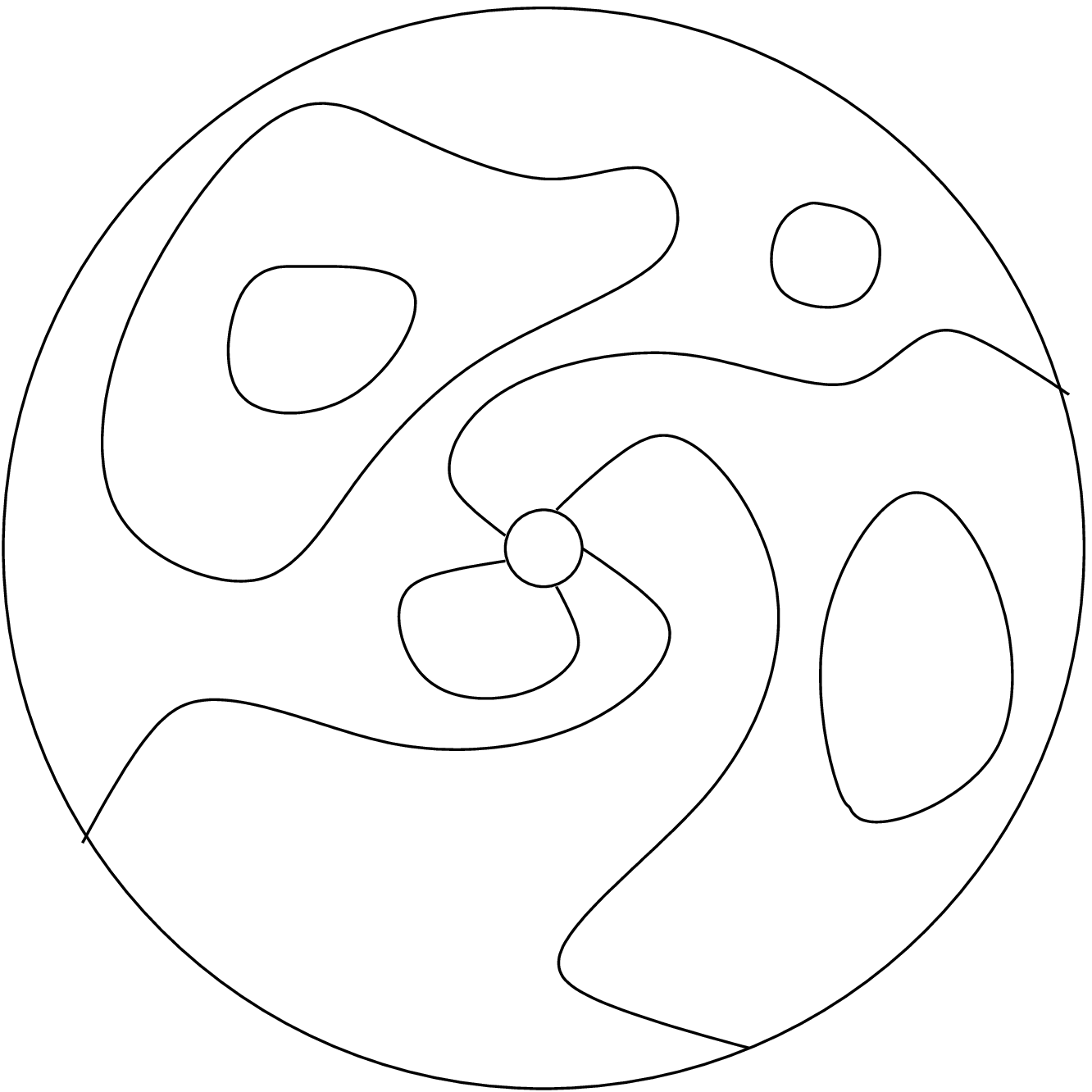}}
\caption{The geometrical set-up. $N$ (here $=3$) open curves connect the boundaries at
$r=\epsilon$ and $R$ of the annulus, intersecting the outer boundary at points
$\{Re^{i\theta_j}\}$. No other open curves are allowed to intersect $r=R$, but
they may intersect $r=\epsilon$, as well as there being any number of
closed loops (except when they carry zero weight, as for $n=0$.)}
\end{figure}
\begin{figure}
\label{fig2}
\centerline{
\epsfxsize=10cm
\epsfbox{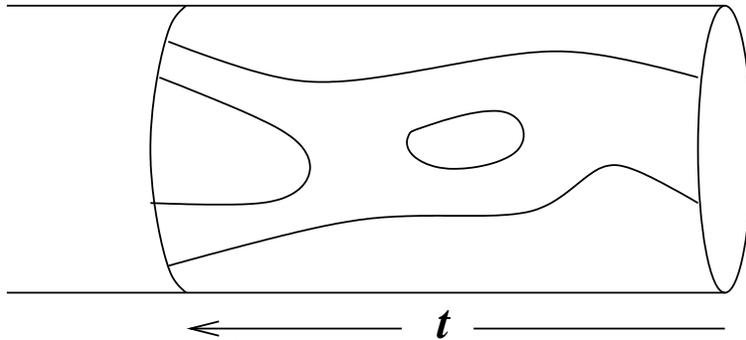}}
\caption{The transfer matrix of Ref.~15 on the cylinder. 
It keeps track of the positions of points where curves intersect the given 
time-slice, as well as their connectivity in the past (but not the future.)
We distinguish between
those points connected to the boundary at $t=0$ ($r=R$), and those which close
for times $t>0$.} 
\end{figure}

\end{document}